\documentclass[10pt,letterpaper]{article}
\usepackage{float}
\usepackage[top=0.85in,left=2.75in,footskip=0.75in]{geometry}
\usepackage{graphicx}
\usepackage{subfigure} 
\usepackage{amsmath,amssymb}

\usepackage{changepage}

\usepackage{textcomp,marvosym}

\usepackage{cite}

\usepackage{nameref,hyperref}

\usepackage[right]{lineno}

\usepackage[nopatch=eqnum]{microtype}
\DisableLigatures[f]{encoding = *, family = * }

\usepackage[table]{xcolor}

\usepackage{array}

\newcolumntype{+}{!{\vrule width 2pt}}

\newlength\savedwidth

\newcommand\thickhline{\noalign{\global\savedwidth\arrayrulewidth\global\arrayrulewidth 2pt}%
\hline
\noalign{\global\arrayrulewidth\savedwidth}}

\raggedright
\usepackage[aboveskip=1pt,labelfont=bf,labelsep=period,justification=raggedright,singlelinecheck=off]{caption}

\makeatletter
\renewcommand{\@biblabel}[1]{\quad#1.}
\makeatother

\usepackage{multirow}

\usepackage{lastpage,fancyhdr,graphicx}
\usepackage{epstopdf}
\fancyheadoffset[L]{2.25in}
\fancyfootoffset[L]{2.25in}
\usepackage{changes}  
\definechangesauthor[name={Reviewer 1}, color=red]{Reviewer 1} 
\definechangesauthor[name={Reviewer 2}, color=blue]{Reviewer 2} 

\reversemarginpar
\begin{document}
\vspace*{0.2in}

\begin{flushleft}
{\Large
\textbf\newline{A transfer learning approach for automatic conflicts detection in software requirement sentence pairs based on dual encoders} 
}
\newline
\\
Yizheng Wang\textsuperscript{1},
Tao Jiang\textsuperscript{1*},
Jinyan Bai\textsuperscript{1},
Zhengbin Zou\textsuperscript{1},
Tiancheng Xue\textsuperscript{1},
Nan Zhang\textsuperscript{1},
Jie Luan\textsuperscript{1}
\\
\bigskip
\textbf{1} School of Mathematics and Computer Science, Yunnan Minzu University, Kunming, China
\\
\bigskip

* jtzwyinternet73@163.com

\end{flushleft}

\clearpage

\section*{Abstract}
Software Requirement Document (RD) typically contains tens of thousands of individual requirements, and ensuring consistency among these requirements is a critical prerequisite for the success of software engineering projects. Automated detection methods can significantly enhance efficiency and reduce costs; however, existing approaches still face several challenges, including low detection accuracy on imbalanced data, limited semantic extraction due to the use of a single encoder, and poor performance in cross-domain transfer learning. To address these issues, this paper proposes a Transferable Software Requirement Conflicts Detection Framework based on SBERT and SimSCE, termed TSRCDF-SS. First, the framework employs two independent encoders—Sentence-BERT (SBERT) and Simple Contrastive Sentence Embedding(SimCSE)to generate sentence embeddings for requirement pairs, followed by a six-element concatenation strategy. Furthermore, the classifier is enhanced by incorporating a two-layer fully connected, alongside a hybrid loss function optimization strategy for feedforward neural network(FFNN) that integrates a variant of Focal Loss, domain-specific constraints, and a confidence-based penalty term. Finally, the framework synergistically integrates sequential and cross-domain transfer learning. Experimental results demonstrate that our framework achieves a 10.4\% improvement in both macro-F1 and weighted-F1 scores in domain settings, and a 11.4\% increase in macro-F1 in cross-domain scenarios.

\section*{1 Introduction}
Requirements Engineering (RE) is a core activity in software development, serving as the foundation for communication between developers, clients, and organizations~\cite{bib1,bib2}. Its primary responsibilities include documenting, identifying, analyzing, and managing requirements~\cite{bib3}, which are commonly presented in the form of Requirement Specification (RS) documents. Due to the widespread applicability of natural language (NL) in RE~\cite{bib4,bib5,bib6}, RS documents are usually written in NL and generally adopt standardized specification expression templates. This pattern has penetrated into various development projects across industries~\cite{bib7,bib8,bib9}. However, RS documents often suffer from issues such as requirement conflicts and redundancy, which pose significant challenges to the development and deployment of software systems. Specifically, a requirement conflict refers to a negative constraint relationship between two requirements, while requirement redundancy indicates that different formulations essentially refer to the same objective~\cite{bib10,bib11}. For example:

R1: The software is compiled from source code using a Java compiler.
R2: The program is executed by a Python interpreter on the server. 

R1 and R2 both concern the execution of code, but the implementation of R1 may hinder the realization of R2, and vice versa. Hence, a conflict exists between R1 and R2.

R3:The software is tested for quality assurance using automated tools.
R4: Quality analysts use automated testing tools for software quality assurance. 

R3 and R4 both require the use of automated tools for software testing. Implementing either requirement inherently satisfies the other, indicating a redundancy between R3 and R4.

Such defects increase project complexity and introduce obstacles to team collaboration. Therefore, effective management of requirements within the documentation and the definition of a complete, unambiguous, and conflict-free target system are of paramount importance~\cite{bib12,bib13}. The industry currently faces a dual dilemma: traditional manual inspection methods suffer from low efficiency and high error rates~\cite{bib10}, while existing automated techniques are often constrained by rigid formatting requirements or dependencies on additional components, limiting their practical applicability. This situation underscores the urgent need for novel requirement management solutions.

Key challenges in software requirement conflict detection include:
\begin{itemize}
	\item Imbalanced data distribution: The inherent sparsity of requirement conflicts and redundancies in real-world scenarios leads to a highly skewed class distribution, making it difficult for models to effectively learn the features of minority classes and thus compromising their generalization capability~\cite{bib11}.
	\item Complex semantics and associative parsing barriers: Requirement texts often involve domain-specific terminology and complex semantic structures. Accurate conflict detection requires not only understanding explicit relationships between requirements but also reasoning about implicit connections based on contextual and domain knowledge~\cite{bib10}.
	\item Heterogeneity in expression: The diverse ways in which stakeholders articulate requirements, such as non-technical vocabulary or linguistic errors and so on, further complicate conflict detection~\cite{bib14}.
\end{itemize}

To address these challenges, this paper proposes an automated detection framework based on a dual-encoder architecture combined with transfer learning strategies. First, the solution is based on the Transformer pre-trained language model BERT and SimCSE, and uses two independent encoders to vectorize and concatenate the demand sentence pairs. Additionally, the classifier is enhanced using a two-layer fully connected network and optimized via a hybrid loss function that integrates a variant of Focal Loss, domain-specific constraints, and a confidence penalty term. Finally, sequential and cross-domain transfer learning is synergistically integrated into the framework.

The main contributions of this paper are as follows:

\begin{enumerate}
	\item{This paper proposes a dual-encoder framework based on SBERT~\cite{bib15} and SimCSE~\cite{bib16}, which utilizes two independent encoders to generate vector representations of requirement sentence pairs. A six-element concatenation strategy is employed to construct a feature representation system with multi-level semantic understanding capabilities, thereby enhancing the model's sensitivity to semantic associations and its expressiveness in capturing conflicting relations.}
	\item{We enhance the FFNN classifier by adopting a two-layer fully connected architecture. This two-layer fully connected design allows for progressive feature abstraction while maintaining moderate model complexity, achieving a balance between performance and efficiency in mid-scale tasks and facilitating richer semantic feature extraction.}
	\item{We introduce a hybrid loss optimization strategy tailored for FFNN. By integrating a variant of Focal Loss, domain-specific constraints, and a confidence-based penalty term, the hybrid loss function dynamically adjusts the focusing parameters. This approach effectively addresses the limitations of traditional cross-entropy loss in handling hard samples, class imbalance, and overfitting.}
        \item{We construct a training mechanism that combines sequential and cross-domain transfer, taking into account the transfer efficiency of pre-training knowledge and the adaptability of target tasks, and improving the generalization performance of the model in heterogeneous domains and task change scenarios.}
\end{enumerate}

The structure of this paper is organized as follows: Section 2 reviews the related work. Section 3 presents the methodology adopted in this paper. Section 4 describes both the in-domain and cross-domain transfer experiments along with the corresponding results. Section 5 discusses potential threats to validity. Finally, Section 6 concludes the paper.

\section*{2 Related Works}
Since the 1980s, numerous researchers have explored innovative methods and technologies to tackle the complexity and variability of NL in RE. For instance, Abbott~\cite{bib17} extracted features from textual requirements based on syntactic patterns, while Aguilera and Berry~\cite{bib18} and Rolland and Proix~\cite{bib19} processed textual relationships in RS documents by identifying words, phrases, and semantic structures within sentences. Since the late 20th century, NLP technologies have been increasingly adopted in the RE domain~\cite{bib4}, with researchers continually introducing novel approaches for a variety of RE tasks. 

\subsection*{2.1 Conflict detection}
Zhao et al.~\cite{bib4} conducted a survey of over 400 related studies and found that the majority of emerging NLP technologies and tools lack practical applicability. The research revealed a significant gap between recent advancements in NLP for RE (NLP4RE) and their actual deployment. Fischbach et al.~\cite{bib20} , through the analysis of various embedded causal relationships, argued that automatically extracting causality from requirements facilitates semantic comparison. Linzen and Baroni~\cite{bib21} employed deep learning techniques for NLP tasks. They built a new causal relationship extraction method based on the NLP architecture of tree recursive neural network, and used it as a basis to detect requirements.

Guo et al.~\cite{bib10} introduced FSARC, a fine-grained semantic analysis-based conflict detector. They categorized contradictions into three main types and seven subcategories, and used heuristic rules and algorithms to identify semantic elements (octuples) extracted via NLP techniques. However, their work did not address how to effectively identify conflicts based on these elements. Zhao et al.~\cite{bib22} pointed out that most research integrating NLP and RE relies heavily on laboratory validation or example-based evaluation, with a notable absence of industrial-scale empirical studies. Tian et al. generated difficult-to-detect adversarial samples through multi-label perturbation of LESSON and saliency guidance of EVADE, providing a reference for the construction of difficult examples and focusing strategies in demand conflict detection\cite{bib44,bib45}. Importantly, existing research has yet to establish a machine learning classification framework specifically for requirement contradictions. Current achievements are largely concentrated on foundational classification tasks such as requirement type identification (e.g. distinguishing functional and non-functional requirements, security requirement detection)~\cite{bib23,bib24}.

Gärtner et al.~\cite{bib25} investigated contradictions in RE environments, offering definitions, taxonomies, and identification methods, and developed a standardized semi-automated solution. Their method enables reviewers to identify contradictions related to the Law of Non-Contradiction without requiring deep familiarity with the RS document context. Subsequently, Gärtner et al.~\cite{bib26} further explored the detection of conditional sentences in RD, analyzing two NLP techniques for identifying conditional expressions and their role in RE. Their findings underscore the importance of conditional constructs in both requirement analysis and conflict detection. Building upon this, Gärtner et al.~\cite{bib27} proposed ALICE (Automatic Logic for Identifying Contradictions in Engineering), which combines formal logic with large language models (LLMs) to detect engineering contradictions in RD. Despite the system's high theoretical and practical value, it exhibits certain limitations: LLMs may generate false positives due to misuse of antonyms, and formal logic-based methods are vulnerable to disruption from filler words or punctuation errors. Additionally, the use of passive voice may further reduce contradiction detection accuracy.

Malik et al.~\cite{bib28,bib46}introduced SR-BERT, a deep learning framework based on Transformer architecture, designed for the automated detection of requirement conflicts and redundancies. SR-BERT can efficiently process the semantic relationship between requirement sentence pairs, thereby identifying conflicts and repetitions, while improving the adaptability and performance of the model through multi-stage refinement training. However, it fails to effectively solve the problem of uneven distribution of dataset labels, especially when conflicting and repeated demand sentence pairs are relatively scarce, the model may be affected by unbalanced data, resulting in overfitting and reduced accuracy.

\subsection*{2.2 Transfer learning}
Transfer learning is a pivotal branch of machine learning that has attracted considerable attention in recent years. Its core principle is to transfer knowledge from one domain to a related but different domain in order to enhance model performance on new tasks. Pan and Yang~\cite{bib29} provided a systematic taxonomy of transfer learning methods, classifying them into instance-based, feature-based, parameter-based, and relation-based frameworks, thereby laying a theoretical foundation for subsequent work. In the NLP domain, transfer learning has markedly enhanced generalization and performance through large-scale pre‑training and cross‑domain knowledge transfer, overcoming challenges posed by data scarcity, domain discrepancies, and high computational costs. Numerous methods leveraging pre-trained language models have surpassed previous state‑of‑the‑art benchmarks~\cite{bib30,bib31}. The success of AlexNet in the 2012 ImageNet competition not only revolutionized computer vision but also catalyzed the widespread adoption of transfer learning. Subsequent pre‑trained architectures—such as VGG, ResNet, and Inception—trained on massive datasets have been fine‑tuned for downstream tasks, yielding substantial gains even in low‑resource settings.

Yosinski et al.~\cite{bib32} empirically demonstrated that lower‑layer representations exhibit higher transferability across tasks, whereas higher‑layer features tend to be more task‑specific. Pan and Yang’s review~\cite{bib29} further clarified the definitions of domain and task in transfer learning, and outlined core concepts and classification schemes, though it did not deeply explore the integration of deep learning with transfer learning nor provide extensive empirical validation on NLP tasks. In 2018, Howard and Ruder~\cite{bib33} introduced the ULMFiT framework, achieving efficient transfer learning in NLP by employing an LSTM‑based pre‑trained language model. ULMFiT’s three‑stage strategy—general‑domain pre‑training, task‑specific fine‑tuning, and classifier fine‑tuning. Its core contributions lie in the discriminative fine-tuning and gradual unfreezing techniques to balance feature updating and alleviate catastrophic forgetting. However, its fine‑tuning hyperparameters lacked systematic optimization and thorough generalization analysis. Building on these advances, Malik et al.~\cite{bib28} designed the SR‑BERT framework using a sequential transfer learning approach. Starting from a general pre‑trained model, SR‑BERT was first fine‑tuned on a large-scale sentence‑pair dataset to obtain an intermediate checkpoint, and then further fine‑tuned on domain‑specific software requirement pairs, enabling the model to capture specific patterns of conflicts and duplicates within the software requirements. Although this transfer learning approach demonstrates excellent performance on non-cross-domain and balanced datasets, it still requires further improvement in scenarios characterized by data scarcity or substantial domain divergence.

\section*{3 Methodology}
\subsection*{3.1 Framework}
\begin{figure}[!h]
\centering 
\includegraphics[width=0.8\textwidth]{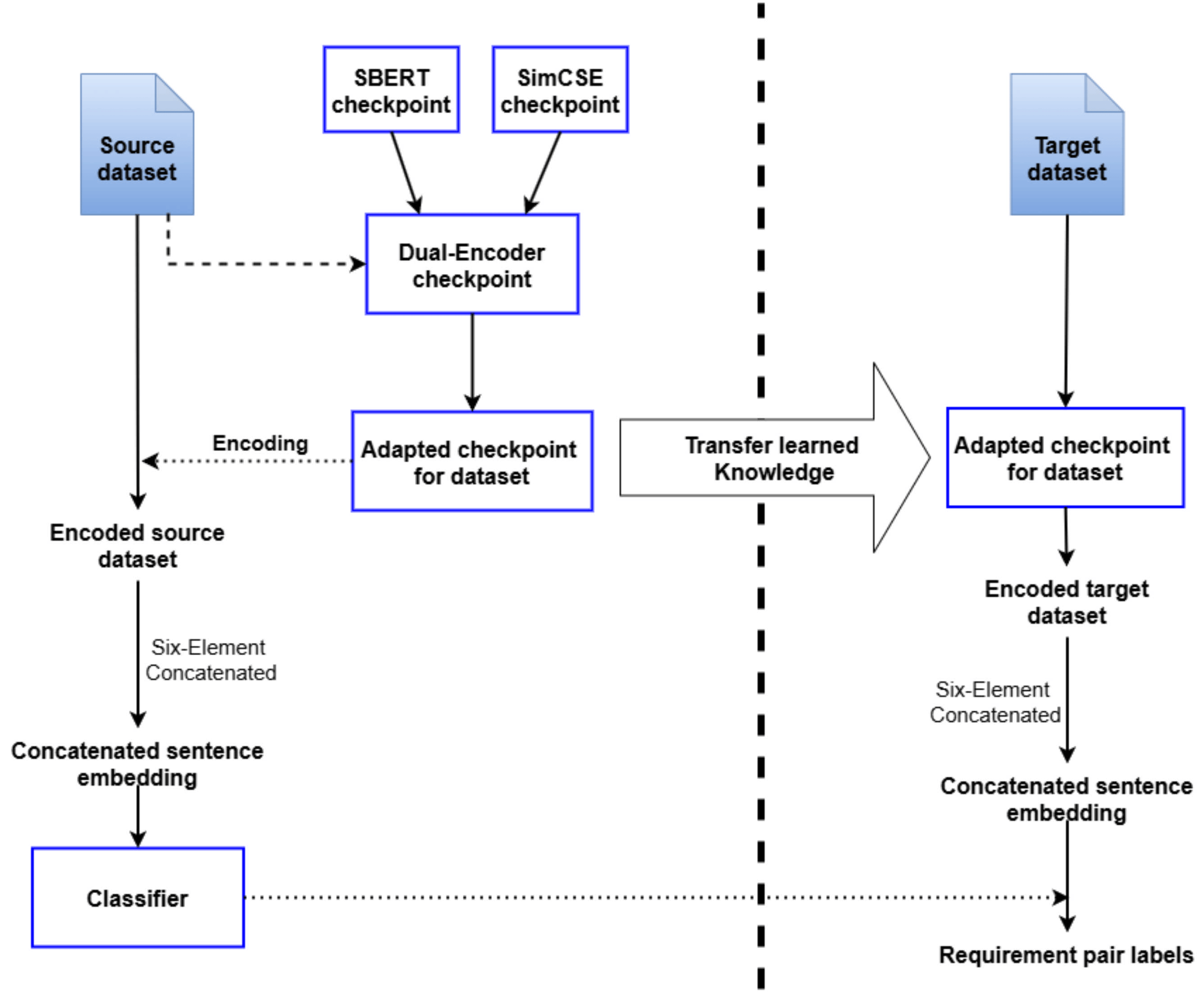} 
\caption{{\bf TSRCDF-SS structure diagram.}
The TSRCDF-SS structure diagram includes a dual encoder, and the improved classifier combines sequential transfer and cross-domain transfer.}
\label{fig1}
\end{figure}

In the task of requirement sentence pair detection, significant challenges persist, such as severely imbalanced label distributions within datasets and the limitations of single encoders in accurately capturing subtle semantic differences or complex logical relationships between sentences, and so forth. To improve detection accuracy, this paper extends the SR‑BERT model by proposing an enhanced framework named TSRCDF-SS. The overall architecture is shown in Fig~\ref{fig1}. First, the framework encodes requirement sentence pairs using two independent encoders, followed by a cross-model hierarchical six-element concatenation strategy to fuse the two semantic representations. This design enables a complementary integration of semantic knowledge by fully leveraging the distinct encoding outcomes.Second, the FFNN classifier adopts a two-layer nonlinear transformation architecture, and a hybrid loss function is introduced, which incorporates a variant of Focal Loss, domain-specific constraints, and a confidence-based penalty term. The two-level fully connected facilitates deeper semantic feature extraction, while the hybrid loss enables more comprehensive optimization of the class probability distribution.Finally, during the model transfer phase, we integrate sequential transfer learning with cross-domain adaptation. The encoder and classifier trained in the source domain are transferred to the target domain to perform conflict prediction on target data. This combined strategy equips the model with enhanced adaptability and fault tolerance in complex scenarios involving both task evolution and domain shifts.

\subsection*{3.2 Build dual encoder}
Sentence embedding techniques encode semantic information into fixed‑dimensional vectors to enhance the efficiency of NLP tasks. In the context of conflict detection, they offer advantages such as reduced computational cost, support for cross‑domain transfer, and facilitation of model collaboration. However, mainstream encoders exhibit limited capacity to represent complex logical phenomena—such as semantic inversion—and the compression inherent in fixed‑size vector spaces can result in the loss of fine‑grained conflict features in long or syntactically complex sentences. Moreover, pre‑training paradigms based on general corpora (e.g. BERT, SBERT) encounter domain‑adaptation bottlenecks. Consequently, this approach still risks accuracy degradation when attempting to balance semantic representation strength against computational efficiency.

Compared to GPT‑2~\cite{bib34}, FastText~\cite{bib35}, and Universal Sentence Encode (USE)~\cite{bib36}, SBERT and SimCSE offer significant benefits for sentence embedding tasks, including more precise semantic representations, higher computational efficiency, and broader applicability. GPT‑2 is primarily optimized for text generation, whereas traditional word‑embedding models (e.g. FastText, USE) neglect contextual semantics and thus struggle to capture logical contradictions in requirement pairs. Therefore, SBERT and SimCSE are better suited to scenarios with stringent demands on embedding quality.

\begin{figure}[!h]
\centering 
\includegraphics[width=0.8\textwidth]{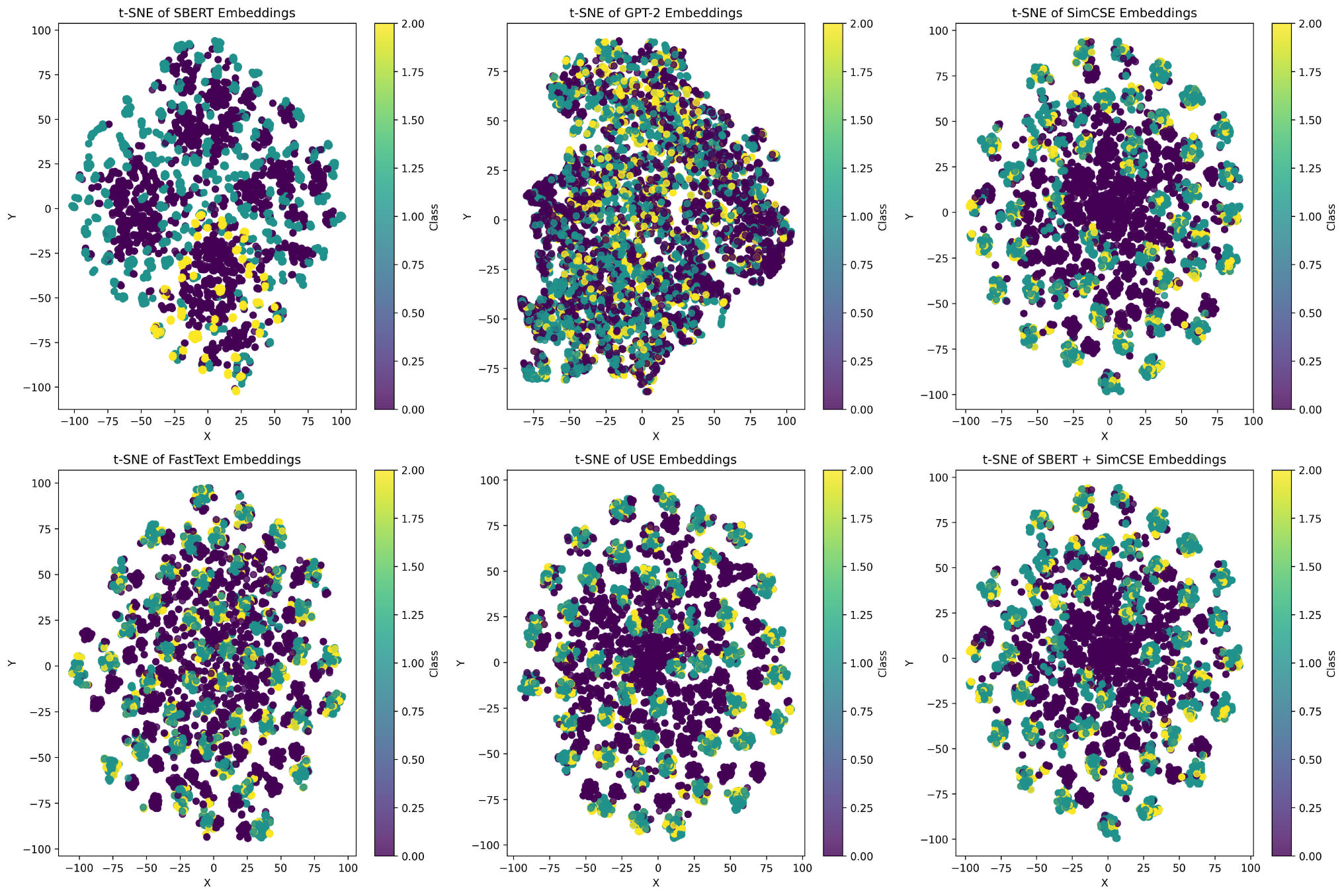}
\caption{{\bf Encoder t-SNE dimensionality reduction projection.}
Comparison of sentence embedding performance of different encoders and combined encoders using t-SNE dimensionality reduction projection. This visualization highlights the differences in encoding capabilities of different encoders. The dataset used is TRAINNLI.}
\label{fig2}
\end{figure}

As shown in Fig~\ref{fig2}, the t-SNE dimensionality reduction reveals that the embeddings generated by SBERT form relatively tight local clusters in the projected space, indicating that its dual-tower siamese architecture effectively captures fine-grained semantic features, however, the boundaries between categories remain blurred, leading to some degree of overlap. FastText, which relies on statistical information at the subword level, tends to cluster morphologically similar but semantically unrelated words together, resulting in disordered clustering structures. The embeddings generated by GPT-2 exhibit strong domain-specific characteristics, which constrain their cross-domain generalization capabilities. SimCSE demonstrates an initial tendency toward cluster formation, with certain categories showing promising intra-cluster cohesion, but the boundaries between data points are not clear . The embeddings from USE display a mild layered structure, but the inter-class boundaries remain indistinct.

In contrast, the SBERT + SimCSE fusion model demonstrates the most favorable visualization performance. The embeddings generated by this model form distinct multi-cluster structures in the projected space, with reduced noise and more concentrated data coverage. Clear boundaries are observed between different categories, with most intra-class samples successfully clustered and inter-class samples well separated. These results indicate that the fusion strategy effectively integrates SBERT’s robust semantic modeling capabilities with the contrastive learning strengths of SimCSE, resulting in more discriminative and structurally coherent semantic representations. Overall, the fusion approach exhibits superior quality in textual embeddings, highlighting its strong potential for application in requirement semantics modeling.

\begin{figure}[!h]
\centering 
\includegraphics[width=0.8\textwidth]{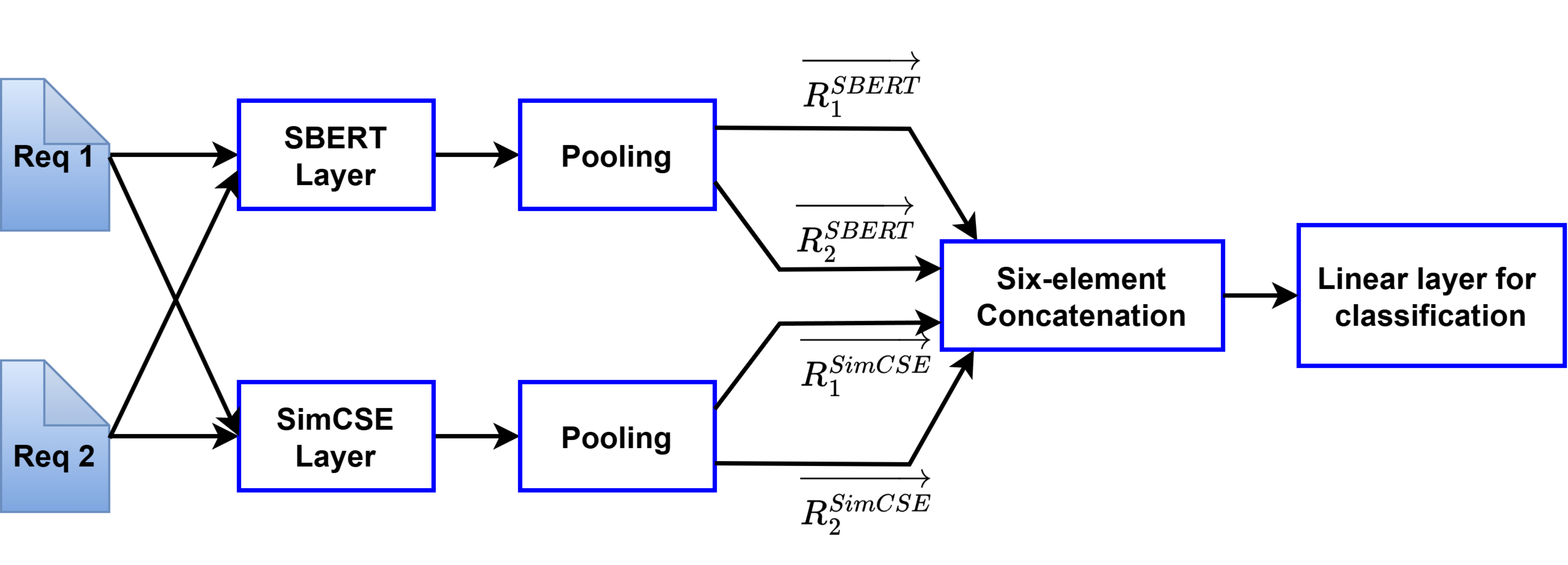}
\caption{{\bf Software requirement pair encoding model based on SBERT and SimCSE.}
Use the SBERT model and SimCSE model to encode individual requirements to obtain their respective embeddings. Then these embeddings are fused to finally obtain the six-element concatenated embedding result.}
\label{fig3}
\end{figure}

Therefore, this paper proposes a fusion sentence embedding framework based on a dual encoder architecture, as illustrated in Fig~\ref{fig3}. Specifically, during the encoding phase, the dual-channel structure enables parallel processing with isolated parameters to independently extract heterogeneous knowledge representations. SBERT, a discriminative model optimized from the BERT architecture, leverages pre-trained sentence-level relational knowledge to generate high-resolution semantic vectors. Its static encoding capabilities are well-suited for capturing fine-grained semantic features in software requirement texts, and achieve efficient alignment of the semantic space through the contrastive learning strategy of the siamese network. Simultaneously, SimCSE—representing the contrastive learning paradigm—employs an unsupervised dropout-based noise perturbation strategy to align positive samples and separate negative ones in latent space. This allows the generated embeddings to possess stronger resistance to contextual noise and improved inter-class discrimination. In the feature fusion stage, a cross-model hierarchical six-element concatenation strategy is employed to achieve collaborative optimization of the heterogeneous representations. Improve upon the three-element concatenation approach of the single-encoder SR-BERT framework proposed in reference~\cite{bib28}, the static semantic vector generated by SBERT and the dynamic context vector of SimCSE are six-element concatenated using the Eq~(\ref{eq:1}). This construction creates a feature space that is orthogonally complementary. The concatenated vectors are fed into a multilayer perceptron classifier, where nonlinear transformations capture higher order semantic associations and produce a conflict probability distribution. Throughout this process, the sextuple concatenation not only enhances sensitivity to subtle semantic shifts but also facilitates discrimination between neutral and conflicting classes. The SBERT vector exhibits a Spearman correlation of 82\% in the semantic similarity calculation task (STS-B benchmark)~\cite{bib15}, while the unsupervised SimCSE improves the previous best average Spearman correlation by 4.2\%, and the supervised SimCSE improves the best average Spearman correlation by 1.24\%~\cite{bib16}. This shows that SimCSE can effectively improve the quality of clustering results when dealing with a small amount of annotated data.

\begin{eqnarray}
\label{eq:1}
    &\overrightarrow{R_{1}^{\text{SBERT}}} \oplus \overrightarrow{R_{2}^{\text{SBERT}}} \oplus \left(\overrightarrow{R_{1}^{\text{SBERT}}} - \overrightarrow{R_{2}^{\text{SBERT}}} \right) \notag \\
    &\quad \oplus \overrightarrow{R_{1}^{\text{SimCSE}}} \oplus \overrightarrow{R_{2}^{\text{SimCSE}}} \oplus \left(\overrightarrow{R_{1}^{\text{SimCSE}}} - \overrightarrow{R_{2}^{\text{SimCSE}}} \right)
\end{eqnarray}

\subsection*{3.3 Improve the classifier and optimize the loss function}
This paper improves on the SR-BERT framework classifier in reference~\cite{bib28}, and introduces a sentence pair classification model based on a two-layer fully connected FFNN. While a single fully connected layer offers structural simplicity and fewer parameters, its nonlinear expressive capacity is relatively limited, making it inadequate to fully capture complex semantic relationships in sentence pairs. Increasing the number of layers theoretically enhances the model's fitting ability, but excessive depth often leads to gradient vanishing or explosion, parameter explosion, and higher computational complexity, thereby amplifying training difficulty and overfitting risks. To balance these trade-offs, we adopt a two-layer fully connected classification model. The model uses encoder-generated embedding vectors as initial representations and performs nonlinear transformation through two-layer fully connected : The first layer consists of 1,500 hidden units with ReLU activation, effectively expanding the feature space and enhancing representational richness. The second layer contains 1,000 hidden units, also utilizing ReLU, to progressively extract higher-order semantic features. Notably, we introduce asymmetric dropout rates of 0.2 and 0.3 between the two layers. This design, validated through grid search experiments, demonstrates that compared to symmetric dropout strategies, asymmetric dropout improves the F1 score on the validation set and alleviates covariate shift within deeper layers. The final output layer uses a Softmax activation function.

Regarding the loss function, although Binary Cross-Entropy (BCE) and Categorical Cross-Entropy (CCE) are commonly employed in classification tasks, both exhibit shortcomings under challenging conditions. BCE has obvious limitations in complex scenarios such as class imbalance, label noise, gradient saturation, and multi-label classification~\cite{bib37}. Meanwhile, CCE yields weak gradient signals for easily classified samples, limiting the model’s ability to learn from difficult instances. To address this, this paper proposes a hybrid loss function optimization strategy for feedforward neural networks, which integrates a variant of focal loss, domain-specific constraints, and a confidence penalty term, to enhance the model’s ability to handle hard samples and improve the smoothness of the output predictions.

Focal Loss~\cite{bib38} introduces a modulating factor$\left(1-p_t\right)^\gamma$, where $p_t$ denotes the predicted probability of the true class and $\gamma$ is the focusing parameter. This mechanism reduces the contribution of easily classified examples and increases the focus on hard-to-classify instances, thereby improving the model’s learning effectiveness for minority classes. The variant of Focal Loss is defined as shown in Eq~(\ref{eq:2}), where C denotes the number of classes, $y_i$ represents the one-hot encoded ground truth label, $p_i$ is the predicted probability obtained via softmax, $w_i$ indicates the class weight, and $\gamma$ is the focusing parameter controlling the weight of easy examples. The dynamic adjustment of $\gamma$ is calculated using Eq~(\ref{eq:3}), where $\gamma_{base}$ is the initial value, $\eta$ is a modulation factor, and ${\rm Accuracy}_{val}$ denotes the validation accuracy, which can be treated as an externally provided scalar.

\begin{eqnarray}
\label{eq:2}
    \mathcal{L}_{\mathrm{Focal}}=-\sum_{i=1}^{c}w_{i}(1-p_{i})^{\gamma}y_{i}\ln(p_{i})
\end{eqnarray}

\begin{eqnarray}
\label{eq:3}
    \gamma=\gamma_{\mathrm{base}}+\eta\cdot\mathrm{Accuracy}_{\mathrm{val}}
\end{eqnarray}

During training, the model may become overly confident in certain predictions, resulting in low-entropy distributions that typically lead to overfitting. To mitigate this, we introduce a confidence penalty term~\cite{bib39}, computed as shown in Eq~(\ref{eq:4}). This term penalizes overconfident predictions by incorporating the negative entropy of the output distribution, thereby enhancing the model’s generalization on unseen data.

\begin{eqnarray}
\label{eq:4}
    \mathcal{L}_{\mathrm{Conf}}=\sum_{i=1}^{c}p_{i}\ln(p_{i})
\end{eqnarray}

In specific application scenarios, prior domain knowledge may suggest that the matching degree of sentence pairs should follow certain distributional characteristics. To this end, we incorporate a domain-specific constraint term~\cite{bib40}, as expressed in Eq~(\ref{eq:5}). For instance, the Kullback-Leibler (KL) divergence is used to measure the discrepancy between the predicted class distribution and a predefined target distribution $q=[q_1,\ldots,q_C]$, guiding the model to produce outputs aligned with domain-specific priors and enhancing its performance on task-specific objectives. Here, $p_{avg}$ represents the average predicted probability of each category in the current batch.

\begin{eqnarray}
\label{eq:5}
    \mathcal{L}_{\mathrm{Domain}}=D_{\mathrm{KL}}\left(q\parallel p_{\mathrm{avg}}\right)=\sum_{i=1}^{c}q_{i}\ln\frac{q_{i}}{p_{avg,i}}
\end{eqnarray}

By combining the three components above, we define the Adaptive Focal Confidence Loss (AFC Loss) as shown in Eq~(\ref{eq:6}), where $\alpha$, $\beta$, $\lambda$ are weighting coefficients. AFC Loss comprehensively optimizes the sentence pair classification task and contributes to improved overall model performance.

\begin{eqnarray}
\label{eq:6}
\mathcal{L}_{\mathrm{AFC}}=\alpha\cdot\mathcal{L}_{\mathrm{Focal}}+\beta\cdot\mathcal{L}_{\mathrm{Conf}}+\lambda\cdot\mathcal{L}_{\mathrm{Domain}}
\end{eqnarray}

\subsection*{3.4 Fusion of sequential transfer and cross-domain transfer}
Sequential transfer learning primarily focuses on the gradual accumulation and transfer of knowledge across a sequence of tasks, while cross-domain transfer learning aims to bridge the gap in data distribution and feature representation between different domains. If in a practical problem where tasks exhibit both temporal dependencies (i.e., requiring sequential learning) and domain heterogeneity (e.g. varying data acquisition conditions, task contexts, or data modalities), it is essential to design models that simultaneously address both temporal progression and domain disparity in order to effectively solve such problems.

In this paper, we integrate sequential transfer learning with cross-domain transfer learning. On one hand, sequential transfer learning enables the model to progressively accumulate and transfer knowledge from previous tasks, allowing for rapid adaptation and efficient updating when facing a series of new tasks. This approach also mitigates the issue of catastrophic forgetting, thereby maintaining continuity and stability in the model's learned knowledge. On the other hand, cross-domain transfer learning allows the model to overcome discrepancies in data distributions or feature spaces between the source and target domains, facilitating the effective extraction and transfer of rich knowledge from the source domain to the target domain. By combining these two strategies, the proposed approach leverages the advantages of continuous knowledge updating offered by sequential transfer learning, while simultaneously addressing domain heterogeneity through cross-domain transfer learning, ultimately enhancing the model's generalization capability and overall stability.

\begin{figure}[!h]
\centering 
\includegraphics[width=0.8\textwidth]{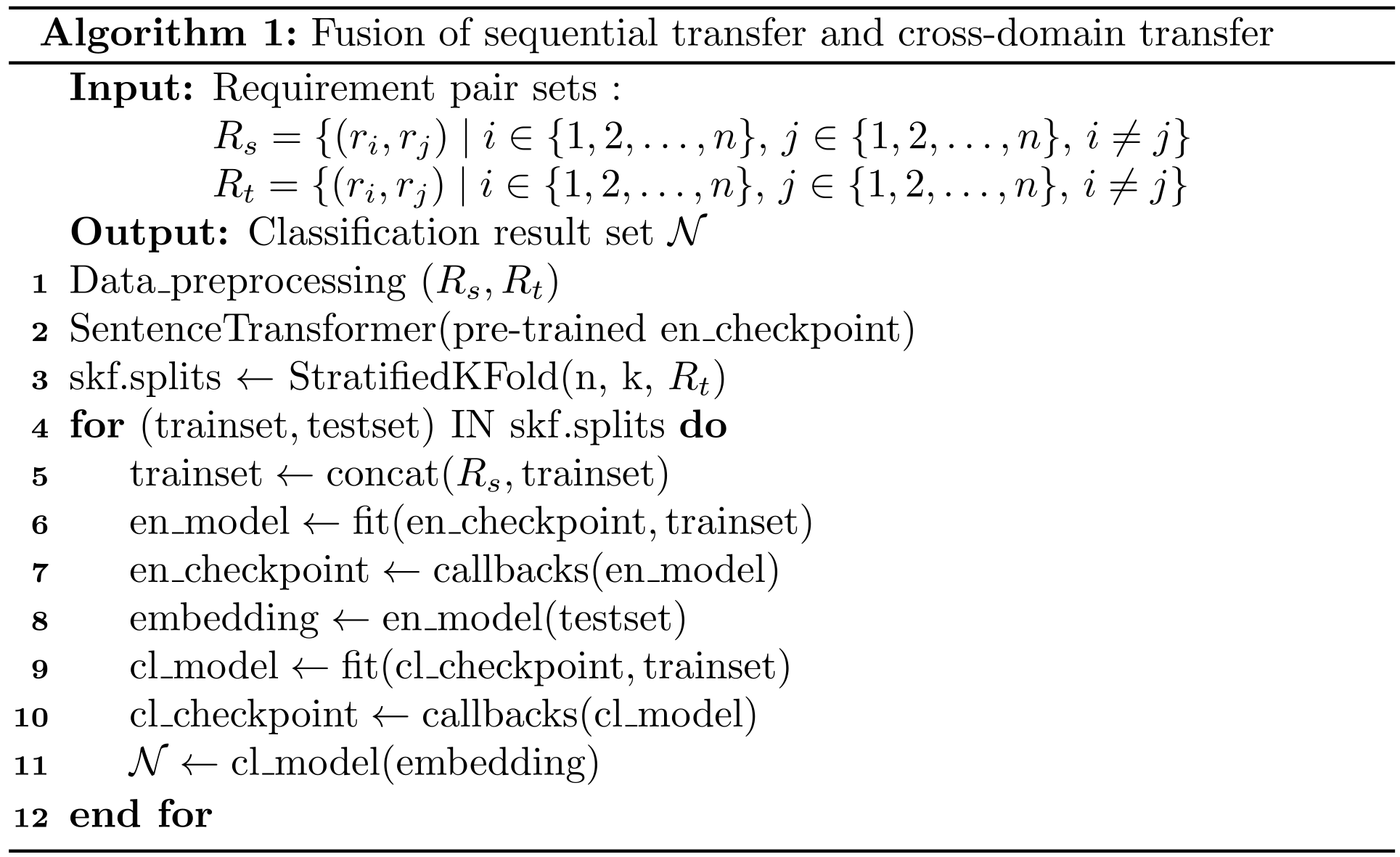}
\caption{{\bf The pseudocode of fusion algorithm of sequential transfer and cross-domain transfer.}}
\label{fig4}
\end{figure}

This paper extends the sequential transfer learning approach presented in reference~\cite{bib28} by proposing a unified process that integrates both sequential transfer learning and cross-domain transfer learning. The algorithm is shown in Fig~\ref{fig4}. The input consists of a source domain requirement pair set and a target domain requirement pair set. First, both sets are preprocessed. Next, a pretrained encoder checkpoint is loaded. In Step 3, the training and testing sets are then divided using n-fold cross-validation. In Step 4, a portion (k/n) of the target domain requirement pairs is combined with the source domain requirement pairs to form a domain-adaptive training set.In Step 5, the encoder is fine-tuned based on the checkpoint and the constructed training set. Step 6 saves the updated encoder checkpoint. Step 7 encodes the testing set using the updated encoder. In Step 8, a classifier is trained on the encoded data, and in Step 9, the classifier checkpoint is saved. Finally, Step 10 outputs the classification results of the testing set.Steps 4 through 10 are repeated for n iterations to complete the cross-validation process.

\section*{4 Experiment}
\subsection*{4.1 Datasets}
This paper integrates multiple publicly available and domain-specific requirement sentence pair datasets in the field of software engineering, encompassing tasks such as general requirement inference, domain-specific requirement analysis, and requirement conflict detection.The datasets used include TRAINNLI, CDN/CN, UAV, WorldVista, PURE, and OPENCOSS, where TRAINNLI and CDN/CN are balanced datasets, and UAV, WorldVista, PURE, and OPENCOSS are imbalanced. The characteristics of each dataset are detailed below:

1. TRAINNLI

This repository provides a specialized Natural Language Inference data set~\cite{bib41}, designed to optimize the performance of the language model for NLP problems in software development. The researchers manually examined various texts from the software development domain and established implicit relationships between different proposals. The texts encompass diverse sources, including software descriptions (e.g. Promise dataset, Pure dataset), user manuals for various software, operating system-related articles (e.g. Windows, official Mac documentation), databases (e.g. MongoDB, official Oracle documentation), cybersecurity (e.g. Mitre documentation), and AWS documentation. The dataset contains 500,000 sentence pairs, manually annotated and balanced across three labels: Entailment, Contradiction, and Neutral.

To align with our research requirements, we preprocessed the dataset by retaining the``gold\_label''``sentence1'' and ``sentence2'' fields. For consistency in the label, the original ``Entailment" and ``Contradiction"  were renamed to ``Duplicate" and ``Conflict," respectively. Given the large scale of the dataset, we partitioned TRAINNLI into smaller subsets for experimental convenience. For instance, TRAINNLI(30000) denotes a randomly sampled subset of 30,000 entries, with similar subsets including TRAINNLI(20000), TRAINNLI(10000)1, TRAINNLI(10000)2, and TRAINNLI(10000)3.

2. CDN/CN

The CDN/CN dataset was curated by Malik et al.~\cite{bib28}, sourced from IBM-DOORS. The CDN dataset includes three categories: Conflict, Duplicate, and Neutral, with Conflict and Duplicate pairs dominating and Neutral pairs limited to half of the total. The CN dataset is a simplified version that excludes Duplicate pairs.

Additional domain-specific datasets used in this paper include: UAV Corpus~\cite{bib42} developed by the University of Notre Dame, which covers requirements related to flight control, sensor integration, and safety protocols for unmanned aerial vehicles. WorldVista EHR Corpus is derived from RDs of medical systems, featuring requirements on patient data management and standardization of clinical workflow. PURE Benchmark Corpus~\cite{bib43} which aggregates requirement sentences from 79 public RDs. The OPENCOSS Corpus sourced from the European Open Platform for Safety Certification project, focuses on embedded system safety certification and is characterized by significant class imbalance. The versions used in the above four datasets are all compiled by the Malik research group~\cite{bib28}.

For balanced datasets, we adopt 5-fold cross-validation to ensure reliable performance estimation. For imbalanced datasets, we use 3-fold cross-validation to mitigate the impact of skewed label distribution during training and evaluation.

\subsection*{4.2 Experimental Environment}
The operating system used was Ubuntu 20.04.1, with Python 3.9.21 as the programming environment. The deep learning framework adopted was PyTorch 1.13.0+cu117. The hardware configuration included an Intel Core i7-14700KF CPU and an NVIDIA GeForce RTX 4070 GPU.

\subsection*{4.3 Experimental Results and analysis}
\subsubsection*{4.3.1 Improved encoder comparison experiment}
\begin{figure}[!h]
\centering 
\includegraphics[width=0.8\textwidth]{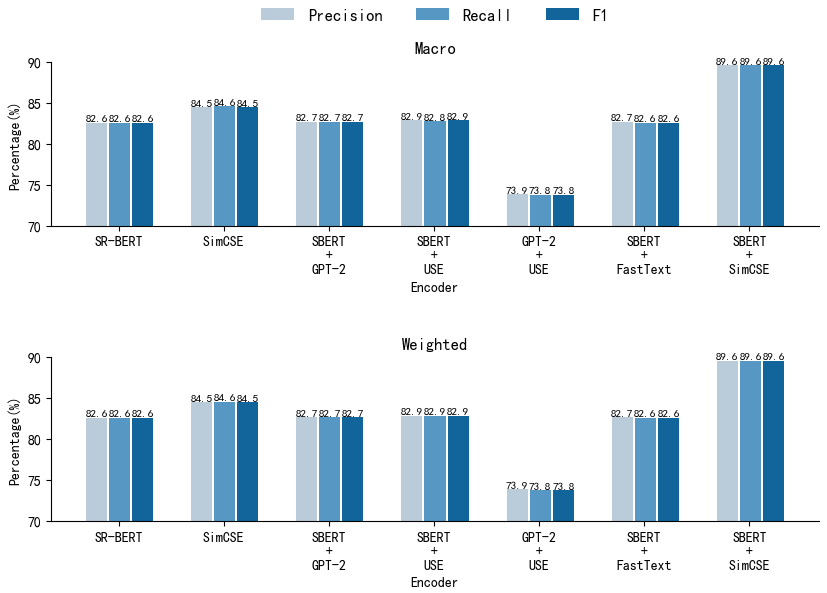}
\caption{{\bf Comparison of encoder combination results.}
 Precision, recall and F1 of different encoder combination experiments on TRAINNLI (30000) dataset.}
\label{fig5}
\end{figure}

In the SR-BERT model, only a single SBERT model was used for sentence embedding. To validate that a dual encoder yields better results for sentence embedding, we conducted experiments on the TRAINNLI (30,000) dataset using various combinations of sentence embedding models. The detailed results are presented in Fig~\ref{fig5}. 

Fig~\ref{fig5} illustrates the performance of SR-BERT and various combined models on the TRAINNLI (30,000) dataset, where the precision, recall, and F1 of the SBERT + SimCSE combination exceed other models, with a weighted F1 score and macro F1 score of 89.6\% and 89.6\%, respectively, which are 7.0\% higher than the weighted F1 score and macro F1 score of SR-BERT. 

\subsubsection*{4.3.2 Improved Classifier comparison experiment}
\begin{figure}[!h]
\centering 
\includegraphics[width=0.8\textwidth]{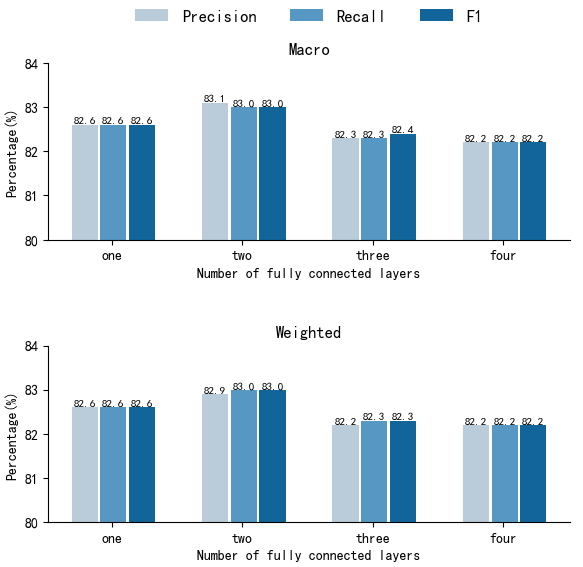}
\caption{{\bf Comparison of results of fully connected layers with different numbers of layers.}
 Results of comparative experiments on different numbers of FFNN layers on the TRAINNLI (30000) dataset.}
\label{fig6}
\end{figure}

In the original SR-BERT model, the classifier composed of a FFNN only had a single fully connected layer. While a single layer has a simple structure and fewer parameters, its ability to express non-linearity is relatively limited, making it difficult to fully capture the complex semantic relationships in sentence pairs. In contrast, using multi-layer fully connected layers allows for the initial extraction of features in the first layer, with subsequent layers further combining and abstracting the non-linear features. This approach improves the model's ability to learn complex patterns. Through experiments, we selected the optimal improvement scheme, and the dataset used was TRAINNLI(30000), with specific results shown in Fig~\ref{fig6}.

Based on the data presented in Fig~\ref{fig6}, we observed that the model with a two-layer fully connected architecture achieved the best performance, with a macro F1 score of 95.0\% and a weighted F1 score of 95.5\%. Through multiple rounds of experimentation and comparison, we found that this two-layer structure strikes an optimal balance between accuracy and stability—it effectively captures semantic features while maintaining high training efficiency and generalization capability.

\begin{figure}[!h]
\centering 
\includegraphics[width=0.8\textwidth]{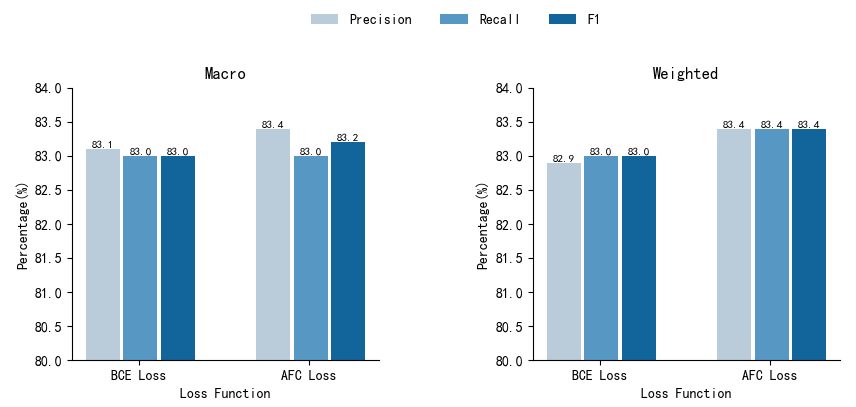}
\caption{{\bf Comparison of results of different loss functions.}
The results of the experiment using a two-layer FFNN and different loss functions on the TRAINNLI (30000) dataset.}
\label{fig7}
\end{figure}

Furthermore, we conducted a comparative experiment on the loss function based on the model of two-layer fully connected layers, using Binary Cross-Entropy and AFC Loss under the same conditions, using the TRAINNLI (30000) dataset. The experimental results are shown in Fig~\ref{fig7}. As illustrated clearly in the bar chart, AFC Loss achieves higher values across all three major metrics—Precision, Recall, and F1-score—with a particularly notable improvement in the F1-score. This validates its applicability and advantage in multi-class mutually exclusive classification tasks, and aligns well with the theoretical analysis.

Experimental results show that the performance with AFC Loss outperforms the performance with Binary Cross-Entropy.

\subsubsection*{4.3.3 Improved encoder and classifier experiment}
In the previous experiments, we separately optimized and compared the encoder and classifier. The goal of this section is to combine both optimized components to evaluate the overall performance improvement on the requirement conflict detection task. Table~\ref{table1} presents the evaluation results from experiments conducted on the TRAINNLI(30000) and CDN/CN datasets. We ran the joint optimized model on these datasets and compared its performance with the independently optimized encoder or classifier models. Overall, the combination of the improved encoder and classifier (Improved Encoder + Improved Classifier) achieved the best results across all datasets. Specifically, on the TRAINNLI dataset, both Macro-F1 and Weighted-F1 reached 91.2\%, which is an improvement of approximately 10.4\% over the baseline SR-BERT (82.6\%). On the CDN dataset, the F1 scores improved to 94.3\%and 96.1\%, respectively. On the CN dataset, the improvement was more limited as the baseline model was already nearing saturation. Furthermore, optimizing the encoder alone (F1 improved from 82.6\% to 89.6\%) was significantly more effective than optimizing the classifier alone (F1 increased to 83.2\%), indicating that high-quality sentence embeddings are crucial for requirement conflict detection. Based on the above analysis, this experiment confirms that the optimization of the encoder contributes the most to performance improvement, while the optimization of the classifier further enhances the model’s discriminative capability. 

\begin{table}[!ht]
\begin{adjustwidth}{-2.25in}{0in} 
\centering
\caption{{\bf The improvement effect of different datasets on the improved model is evaluated, and ablation comparison experiments are performed on the TRAINNLI (30000) dataset (Unit:\%).}}
\begin{tabular}{|l|l|c|c|c|c|c|c|}
\hline
\multirow{2}{*}{\bf Dataset} & \multirow{2}{*}{\bf Model} & \multicolumn{3}{c|}{\bf Macro} & \multicolumn{3}{c|}{\bf Weighted} \\
\cline{3-8}
& & Precision & Recall & F1 & Precision & Recall & F1 \\ \thickhline
\hline
\multirow{4}{*}{\shortstack{\bf TRAINNLI\\(30000)}} 
& SR-BERT & 82.6 & 82.6 & 82.6 & 82.6 & 82.6 & 82.6 \\
& Improved encoder + basic classifier & 89.6 & 89.6 & 89.6 & 89.6 & 89.6 & 89.6 \\
& Basic encoder + improved classifier & 83.4 & 83.0 & 83.2 & 83.4 & 83.4 & 83.4 \\
& Improved encoder + improved classifier & 91.4 & 91.0 & 91.2 & 91.2 & 91.2 & 91.2 \\
\hline
\multirow{2}{*}{\bf CDN} 
& SR-BERT & 93.6 & 93.8 & 93.7 & 95.3 & 95.3 & 95.3 \\
& Improved encoder + improved classifier & 94.2 & 94.4 & 94.3 & 96.1 & 96.0 & 96.1 \\
\hline
\multirow{2}{*}{\bf CN} 
& SR-BERT & 99.4 & 99.1 & 99.2 & 98.9 & 99.6 & 99.4 \\
& Improved encoder + improved classifier & 99.4 & 99.2 & 99.3 & 99.3 & 99.5 & 99.4 \\
\hline
\end{tabular}
\label{table1}
\end{adjustwidth}
\end{table}

Table~\ref{table2} compares our model with other models on the TRAINNLI(30000) dataset in a non-cross-domain scenario. The combination of the improved encoder and classifier performed best, with an F1 score of 91.2\%, which is about a 5\% improvement over BERT-MNLI (86.3\%), demonstrating strong overall discriminative power and balanced performance across classes. Compared to traditional shallow models such as TF-IDF + SVM and Okapi BM25 + MLP, our framework exhibits clear advantages in semantic modeling and feature abstraction, indicating that sentence‑level pre-trained models are better suited to the requirement conflict detection task. These experimental results fully validate the effectiveness of the proposed model design for requirement conflict detection.

\begin{table}[!ht]
\begin{adjustwidth}{-2.25in}{0in} 
\centering
\caption{
{\bf Performance evaluation results on the TRAINNLI (30000) dataset (Unit:\%).}
}
\begin{tabular}{|l|c|c|c|c|c|c|}
\hline
\multirow{2}{*}{\bf Models} & \multicolumn{3}{c|}{\bf Macro} & \multicolumn{3}{c|}{\bf Weighted} \\ \cline{2-7}
& \bf Precision & \bf Recall & \bf F1 & \bf Precision & \bf Recall & \bf F1 \\ \thickhline
TF-IDF+SVM                        & 79.1 & 79.1 & 79.1 & 79.0 & 79.1 & 79.1 \\ \hline
OkapiBM25+MLP                    & 85.7 & 85.7 & 85.7 & 85.7 & 85.7 & 85.7 \\ \hline
SR-BERT                          & 82.6 & 82.6 & 82.6 & 82.6 & 82.6 & 82.6 \\ \hline
Deberta-base-mnli                & 86.1 & 86.1 & 86.1 & 86.1 & 86.1 & 86.1 \\ \hline
Bert-base-uncased-MNLI           & 86.3 & 86.2 & 86.3 & 86.3 & 86.2 & 86.3 \\ \hline
Improved encoder + improved classifier & 91.4 & 91.0 & 91.2 & 91.2 & 91.2 & 91.2 \\ \hline
\end{tabular}
\label{table2}
\end{adjustwidth}
\end{table}

\subsubsection*{4.3.4 Cross-domain transfer experiment}
This section of the experiment utilizes TSRCDF-SS for cross-domain transfer learning on requirement sentence pairs. By leveraging the correlation between the source and target domains, the model's ability to generalize across data from different domains is enhanced, thereby further optimizing the overall performance of requirement conflict detection. Table~\ref{table3} presents the evaluation results for cross-domain training on different dataset combinations.

\begin{table}[!ht]
\begin{adjustwidth}{-2.25in}{0in} 
\centering
\caption{{\bf Evaluation of cross-domain models trained with different combinations of requirement pair datasets (Unit:\%).}}
\begin{tabular}{|l|l|c|c|c|c|c|c|}
\hline
\multirow{2}{*}{\bf Target Data} & \multirow{2}{*}{\bf Source Data} & \multicolumn{3}{c|}{\bf macro} & \multicolumn{3}{c|}{\bf weighted} \\
\cline{3-8}
 & & \bf Precision & \bf Recall & \bf F1 & \bf Precision & \bf Recall & \bf F1 \\ \thickhline
\hline
\multirow{6}{*}{UAV} 
& CN & 99.9 & 64.8 & 72.7 & 99.8 & 99.8 & 99.8 \\
& WorldVista & 99.9 & 62.0 & 69.3 & 99.8 & 99.7 & 99.7 \\
& PURE & 83.2 & 52.8 & 55.0 & 99.7 & 99.7 & 99.6 \\
& OPENCOSS & 99.9 & 63.0 & 70.5 & 99.8 & 99.8 & 99.7 \\
& WorldVista, PURE & 99.9 & 64.8 & 72.8 & 99.8 & 99.8 & 99.8 \\
& WorldVista, PURE, OPENCOSS & 99.9 & 67.6 & 76.0 & 99.8 & 99.8 & 99.8 \\
\hline
\multirow{6}{*}{WorldVista} 
& CN & 99.9 & 63.8 & 71.6 & 99.8 & 99.8 & 99.7 \\
& UAV & 99.9 & 65.2 & 73.3 & 99.8 & 99.8 & 99.7 \\
& PURE & 99.9 & 57.6 & 62.9 & 99.7 & 99.7 & 99.6 \\
& OPENCOSS & 99.9 & 65.7 & 73.9 & 99.8 & 99.8 & 99.7 \\
& UAV, PURE & 99.9 & 66.2 & 74.4 & 99.8 & 99.8 & 99.7 \\
& UAV, WorldVista, OPENCOSS & 99.9 & 66.7 & 74.9 & 99.8 & 99.8 & 99.7 \\
\hline
\multirow{6}{*}{PURE} 
& CN & 99.7 & 63.3 & 70.9 & 99.3 & 99.3 & 99.1 \\
& UAV & 99.6 & 60.0 & 66.1 & 99.3 & 99.3 & 99.0 \\
& WorldVista & 94.7 & 66.6 & 73.9 & 99.3 & 99.4 & 99.2 \\
& OPENCOSS & 99.7 & 65.0 & 72.9 & 99.4 & 99.4 & 99.2 \\
& UAV, WorldVista & 99.7 & 66.7 & 74.8 & 99.4 & 99.4 & 99.2 \\
& UAV, WorldVista, PURE & 99.7 & 67.5 & 75.8 & 99.4 & 99.4 & 99.3 \\
\hline
\multirow{6}{*}{OPENCOSS} 
& CN & 66.6 & 51.7 & 53.0 & 99.8 & 99.9 & 99.8 \\
& UAV & 49.9 & 50.0 & 50.0 & 99.7 & 99.9 & 99.8 \\
& WorldVista & 49.9 & 50.0 & 50.0 & 99.7 & 99.9 & 99.8 \\
& PURE & 66.6 & 51.7 & 53.0 & 99.8 & 99.9 & 99.8 \\
& WorldVista,PURE & 83.3 & 56.7 & 60.7 & 99.8 & 99.9 & 99.8 \\
& UAV,WorldVista,PURE & 92.5 & 66.7 & 73.7 & 99.9 & 99.9 & 99.9 \\
\hline
\multirow{7}{*}{CDN} 
& NLI(10000)1 & 89.2 & 88.7 & 88.7 & 90.7 & 90.2 & 90.3 \\
& NLI(20000) & 88.8 & 87.8 & 88.0 & 90.0 & 89.5 & 89.5 \\
& NLI(30000) & 89.1 & 86.5 & 87.5 & 89.7 & 89.2 & 89.2 \\
& NLI(10000)1,NLI(10000)2 & 87.0 & 86.7 & 86.7 & 89.3 & 88.8 & 88.9 \\
& NLI(10000)1,NLI(10000)2,NLI(10000)3 & 90.0 & 90.0 & 89.8 & 91.5 & 91.2 & 91.2 \\
\hline
\end{tabular}
\label{table3}
\end{adjustwidth}
\end{table}

The results in Table~\ref{table3} indicate that, in binary classification tasks, the weighted metrics are consistently close to 99.8\%, suggesting excellent performance in recognizing the majority class. However, the Recall under the macro average is significantly lower. Compared to the cross-domain experiments conducted by Malik et al., all evaluation metrics show varying degrees of improvement, with a notable 11.4\% increase when the target domain dataset is OPENCOSS. For the three-class classification tasks (TRAINNLI and CDN), the F1 scores demonstrate strong performance, with a balanced distribution across both weighted and macro averages.

In the binary classification task, the low macro recall value is primarily due to the fact that this metric computes a simple average across all classes, without accounting for the disparity in sample sizes between the classes. In imbalanced datasets, the majority class, with ample samples, is likely to achieve a high recall, while the minority class, with fewer samples, tends to have a higher false-negative rate. This imbalance significantly lowers the overall recall in the equal-weighted calculation. In contrast, the weighted avg gives more weight to categories with more samples, maintaining a higher overall performance. The low macro recall highlights the model's weakness in recognizing minority classes.

\section*{5 Discuss}
\subsubsection*{5.1 Threats to Validity}
This paper adopts a dual-encoder architecture and transfer learning strategies to enhance the accuracy and generalization capability of requirement conflict detection tasks. However, there are still potential threats to the effectiveness of the proposed approach:

\begin{enumerate}
	\item{Imbalanced Dataset Impact: Some datasets (e.g. UAV, WorldVista, PURE, OPENCOSS) suffer from class imbalance, resulting in lower recall for the minority classes, which in turn affects the performance of macro metrics. Although weighted loss functions and transfer learning strategies have been employed to solve this issue, the detection of minority class requirement conflicts may still be insufficient.}
	\item{Cross-domain Transfer Generalization: This paper adopts a method that combines sequential transfer learning with cross-domain transfer learning to adapt to demand conflict detection tasks in different fields. However, semantic differences between different domains may limit the effectiveness of feature transfer. For instance, the requirement expressions in the medical domain (WorldVista) and embedded systems (OPENCOSS) differ significantly, which could affect the model's adaptability to new domains.}
	\item{Limitations in Experimental Setup: The experiments were primarily conducted on several well-known publicly available datasets, which may introduce dataset-specific biases and affect the generalizability of the results. Furthermore, due to limitations in experimental environment and computational resources, the paper did not test on large-scale datasets, which may affect the comprehensive evaluation of requirement conflict detection capabilities in real-world industrial environments.}
\end{enumerate}

\subsubsection*{5.2 Limitations and future work}
In addition to the aforementioned threats to validity, the current TSRCDF-SS framework primarily relies on the semantic representation capabilities of pre-trained language models, without incorporating structured knowledge such as domain ontologies, standardized terminologies, or logical rules. As a result, it struggles to accurately identify complex requirements involving conditional statements, reasoning chains, or logical constraints. Moreover, the model’s predictions lack interpretability, making it difficult to provide clear explanations or locate the underlying causes of conflicts, which limits its practicality and reduces user trust. Future research will focus on enhancing the model’s robustness to imbalanced data, introducing large language models to support hybrid reasoning for improved interpretability, integrating domain knowledge and knowledge graphs to deepen semantic modeling, and extending the task to multi-label conflict type recognition for more fine-grained detection. Furthermore, efforts will be made to develop explainable and interactive tool systems to facilitate the practical deployment of this method in real-world software engineering projects.

\section*{6 Conclusion}
This paper proposes an automatic requirement conflict detection framework based on a dual-encoder architecture that integrates SBERT and SimCSE, combined with a transfer learning strategy. Experimental results demonstrate that the framework performs excellently across multiple requirement sentence pair datasets, significantly improving both the accuracy and generalization ability of conflict detection. The dual-encoder structure enhances the quality of sentence embeddings by leveraging the strengths of SBERT and SimCSE, thereby improving the model's capability in capturing semantic relations between requirement pairs. On the TRAINNLI dataset, the F1 scores surpass those of other mainstream models. The two-layer fully connected network architecture, along with a hybrid loss function, enables the model to maintain stable performance in complex classification tasks and reduces the risk of overfitting. Furthermore, the transfer learning strategy effectively improves cross-domain detection by combining sequential and cross-domain transfer learning. This allows the model to maintain high recognition performance even on imbalanced datasets. Overall, the experiments indicate that the proposed method is well-suited for requirement conflict detection tasks across different domains.



\nolinenumbers

%
%
%

\end{document}